\documentclass[10pt,conference]{IEEEtran}
\IEEEoverridecommandlockouts
\usepackage{cite}
\usepackage{amsmath,amssymb,amsfonts}
\usepackage{algorithmic}
\usepackage{graphicx}
\usepackage{textcomp}
\usepackage{xcolor}
\usepackage{tabularx}
\usepackage{booktabs}
\usepackage{makecell}
\usepackage{listings}
\usepackage{caption}
\usepackage[frozencache,cachedir=minty]{minted}
\usepackage[hyphens]{url}
\usepackage{csvsimple}
\usepackage{tikz}
\usepackage{pgfplots}
\pgfplotsset{compat=1.17}
\usepackage{hhline}

\usepackage{siunitx}
\usepackage{todonotes}
\presetkeys%
    {todonotes}%
    {inline}{}

\usepackage[normalem]{ulem}
\def\BibTeX{{\rm B\kern-.05em{\sc i\kern-.025em b}\kern-.08em
    T\kern-.1667em\lower.7ex\hbox{E}\kern-.125emX}}
\begin{document}

\title{Beyond the C: Retargetable Decompilation using Neural Machine Translation}
\author{
    \IEEEauthorblockN{Iman Hosseini}
    \IEEEauthorblockA{\textit{New York University}\\
    iman.hosseini@nyu.edu}
    \and
    \IEEEauthorblockN{Brendan Dolan-Gavitt}
    \IEEEauthorblockA{\textit{New York University}\\
    brendandg@nyu.edu}
}

\IEEEoverridecommandlockouts
\makeatletter\def\@IEEEpubidpullup{6.5\baselineskip}\makeatother
\IEEEpubid{\parbox{\columnwidth}{
    Workshop on Binary Analysis Research (BAR) 2022 \\
    24 April 2022, San Diego, CA, USA \\
    ISBN 1-891562-76-2 \\
    https://dx.doi.org/10.14722/bar.2022.23009 \\
    www.ndss-symposium.org
}
\hspace{\columnsep}\makebox[\columnwidth]{}}

\maketitle

\begin{abstract}
The problem of reversing the compilation process, decompilation, is an important tool in reverse engineering of computer software. Recently, researchers have proposed using techniques from neural machine translation to automate the process in decompilation. Although such techniques hold the promise of targeting a wider range of source and assembly languages, to date they have primarily targeted C code. In this paper we argue that existing neural decompilers have achieved higher accuracy at the cost of requiring language-specific domain knowledge such as tokenizers and parsers to build an abstract syntax tree (AST) for the source language, which increases the overhead of supporting new languages. We explore a different tradeoff that, to the extent possible, treats the assembly and source languages as plain text, and show that this allows us to build a decompiler that is easily \emph{retargetable} to new languages. We evaluate our prototype decompiler, Beyond The C (BTC), on Go, Fortran, OCaml, and C, and examine the impact of parameters such as tokenization and training data selection on the quality of decompilation, finding that it achieves comparable decompilation results to prior work in neural decompilation with significantly less domain knowledge. We will release our training data, trained decompilation models, and code to help encourage future research into language-agnostic decompilation.
\end{abstract}

\begin{IEEEkeywords}
	decompilation; deep learning; transformers
\end{IEEEkeywords}

\section{Introduction}

Decompilation is the problem of reversing the compilation process. It is an important tool in reverse engineering of computer software, to support use cases such as malware analysis, security auditing, and re-engineering of legacy code. But binary analysis in general, and decompilation in particular, are not easy~\cite{noteasy,phoenix,johnny}. As a result, decompilers for binary code are few and far between, and require a great deal of of engineering to produce acceptable output.

The traditional method of decompilation is based on lifting the binary into an architecture-independent intermediate language, and then using control flow recovery, type inference, and other program analyses to recover high level code, typically followed by some rewriting based on common C idioms to improve the readability of the resulting code~\cite{nogoto,phoenix,johnny}. Decompilers created in this way are expensive to develop and, for this reason, all binary decompilers we are aware of target the C programming language. Aside from difficulties with retargetability, traditional decompilers also tend to produce non-idiomatic code that may be difficult to understand (e.g., containing excessive GOTO statements~\cite{nogoto}).

Recently, there has been interest in a new approach to decompilation using Neural Machine Translation (NMT)~\cite{Katz2018,coda,Katz2019TowardsND,nbref}. These methods aim to leverage neural networks to mitigate some of the drawbacks of traditional methods, by treating the decompilation problem as a \emph{translation} task and training on a large number of assembly/source pairs. Although these methods offer the promise of cheaply generating decompilers for diverse languages \emph{automatically}, to date this promise has gone unrealized: the neural decompilers referenced above all target C. We argue that this is, in part, because although at first they appear to be language agnostic, they in fact rely on domain knowledge and tools that may be difficult to come by for arbitrary languages:

\begin{itemize}
    \item Integration with the compiler (e.g., a custom compiler pass or plugin)
    \item Parsing/lexing the high level source code or the assembly (dependence on language or CPU architecture)
    \item Requiring awareness of assembly instruction semantics or types (dependence on CPU architecture)
    \item Lifting the binary to an intermediate language (dependence on lifter and that intermediate language)
    \item Working with Abstract Syntax Trees (dependence on that language/AST format)
\end{itemize}

Although this domain knowledge can improve accuracy, particularly with smaller models, it increases the effort required to support a new language: feature extraction tools need to be adapted or rewritten, the new language may have features that do not map cleanly to the existing model architecture, etc. Prior work in neural decompilation generally follows this path, using domain-specific analysis in an attempt to make the model's job easier.

In this paper we explore a different tradeoff, treating the training data as text and avoiding domain knowledge wherever possible. Rather than attempting to be aware of the syntax of each language, we simply extract pairs of source and assembly at the level of whole functions and use this to train a translation model. This greatly simplifies the task of collecting training data and allows us to easily handle new languages as long as we can collect suitable function pairs. We demonstrate this flexibility by training decompilation models for C and, for the first time, Fortran, Go, and OCaml. For C, we achieve results comparable in quality to prior work~\cite{Katz2018} while retaining the ability to easily adapt to new languages.

We note that the purpose of this paper is \emph{not} to introduce new decompilation or machine learning technique. Rather, we are interested in a) how far we can get using the most recent off-the-shelf techniques from machine learning, b) how little domain knowledge we can get away with while retaining useful decompilation quality, and c) whether our design choices allow neural decompilation to make good on its promise to deliver \emph{retargetable} decompilers for multiple languages. Given the amount of progress made in neural machine translation since the most recent comparable related work (Katz et al.~\cite{Katz2018} from 2018) in decompilation, we expect that advances in machine learning (longer sequence lengths and more sophisticated architectures such as the transformer~\cite{attent}) can free us from many of the language-specific features that were imposed by less capable models.

The main contributions of this work are as follows:
\begin{enumerate}
    \item We investigate the viability of discarding language-specific domain knowledge in neural decompilation by designing a decompilation system that treats code as plain text.
    
    \item We demonstrate retargetability by training decompilation models for four programming languages, including three that have never been the subject of research in decompilation.
    
    \item We provide an evaluation of the models in terms of speed and quality of decompilation, and give cross-language comparisons that illuminate differences between languages relevant to decompilation.
    
    \item We will be making our training corpora, trained models and code available to enable further research in neural decompilation.
\end{enumerate}

\section{Background}

\begin{figure*}[t]
	\centering
	\includegraphics[width=\textwidth]{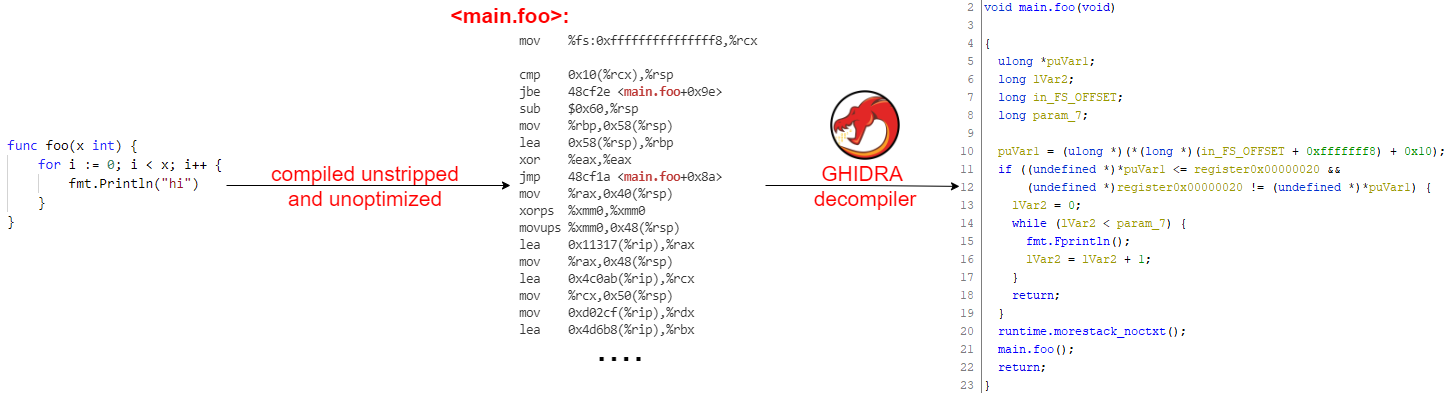}
	\caption{GHIDRA decompilation of a simple Go function}
	\label{ghidrex}
\end{figure*}

\subsection{Challenges of Decompilation}

Decompilation of binary code is a difficult problem. First, as detailed in Meng and Miller~\cite{noteasy}, even basic problems like distinguishing code from data and recovering a control flow graph from a binary executable are hard. The problem of accurately disassembling a binary executable is sometimes known as ``reassembleable disassembly''~\cite{ouroboros}. We will not dwell too much on the problem of low-level disassembly in this paper, however, and will assume that we can recover assembly code in a form similar to what the compiler produces (i.e., with labels intact and data distinguished from code). Because we are primarily interested in \emph{source language retargetability}, we consider in this paper only the familiar (64-bit) x86 architecture, but we expect our techniques to work equally well on other architectures. x86 is also the best-supported architecture for traditional decompilers, allowing more direct comparisons.

Beyond the challenge of disassembly, decompilation from assembly is still hard. Assembly code is much lower level than any modern programming language, and a great deal of information is discarded during the compilation process. For example, variable names and high-level types are lost, and optimizing compilers can entirely omit code if it can be proved to be unreachable (e.g., guarded by an always-false conditional). Compilers are also free to reorder and transform code significantly as long as semantics are preserved, and so can apply transformations like changing integer division into an equivalent multiplication. Such transformations can be very difficult to invert without resorting to hand-crafted patterns specific to each compiler.

Finally, compilers may add significant amounts of code to the program that do not correspond to anything the programmer wrote. This code typically is used to implement runtime support needed by higher-level language features, such as exceptions, bounds checks, etc. If the decompiler is not aware of such automatically generated code, it may include it in the decompiled output, even though it is not helpful to human analysts. These shortcomings can be seen in Figure~\ref{ghidrex}, which shows (from left to right) a small function in Go, a portion of its compiled assembly, and the result of decompilation with Ghidra. It also demonstrates how traditional decompilers struggle at non-C languages. We call this problem of matching low-level assembly to high-level code the \emph{attribution} problem.

\subsection{Neural Machine Translation}

Neural Machine Translation (NMT) is the application of neural networks to translate text from one language to another. NMT systems are widely used for translation of natural languages. Current state-of-the-art NMT systems typically use the transformer architecture~\cite{transf}. SuchAs used in NMT, such models are a composition of an encoder and decoder \cite{cho,s2s}. Input text is \emph{tokenized} into tokens from a fixed vocabulary and fed to the encoder; the decoder emits the output tokens in the target language. The decoder generates, at each step, a probability distribution over the tokens in the target vocabulary, assigning a probability to each token in the vocabulary. By finding the most probable sequence of tokens, a translation can be generated from the model. As the search space for possible sequences is intractably large ($V^N$ where $V$ is size of vocabulary and $N$ is sequence length), greedy methods such as beam search and nucleus (top-$p$) sampling are used to find a solution~\cite{beam,nucleus}.

Text can be decomposed into tokens of varying granularity: tokens can be words, characters or bytes, or anything in between. The internals of a model---i.e., how the encoder and decoder are connected---vary as well  and consist of multiple different layers and connections. The model is trained over a corpus of example translations, where each sample consists of a sentence in the source language and its translation. The loss function used to train the model is a differentiable measure (typically cross-entropy) of how the output probability distribution is from the ground truth. The model is then trained by minimizing the loss function using gradient descent. Neural decompilation is the application of NMT to decompilation, seeing decompilation as translation from assembly to source code.

\subsection{Neural Decompilation}

Researchers have been studying application of NMT methods to decompilation~\cite{Katz2018,Katz2019TowardsND,coda}. These works have focused on the C language and used Recurrent Neural Network (RNN) architectures as their sequence to sequence model. These models have feedback loops (thus called ``Recurrent'') that allow information to flow through the nodes of the encoder and decoder laterally. There is more recent work~\cite{nbref} which uses transformer~\cite{transf} models, a more recent sequence to sequence model which has enjoyed great success in the field of Natural Language Processing (NLP)~\cite{transfsota}. Transformers make a few changes to the RNN architecture, including removing the feedback loops, which allows training to be more efficiently parallelized.

\subsection{Classical vs. Neural Decompilation}

Traditional decompilers apply classic program analyses such as control flow recovery, data flow analysis, etc., and use domain knowledge about the source level language and common compiler idioms to construct a source-level representation. By contrast, neural decompilation treats the problem as one of neural machine translation, and uses deep neural networks to learn a mapping between assembly and source code from many training samples.

Because the neural networks on which they are based are not \emph{interpretable}, neural decompilers are opaque and offer no guarantees about the correctness of the output. This means that when they make mistakes, it can be difficult (or impossible) to diagnose what went wrong or to fix them. In addition, whereas a traditional decompiler can refuse to provide a decompilation when it encounters some construct it cannot handle well, neural decompilers will always provide some output, which may be nonsensical when the input is very different from code in its training set.

These disadvantages are balanced by some unique benefits of NMT-based decompilation. First, because it needs only training data (which can be easily obtained with a compiler) and GPU time, neural decompilation holds the promise of removing the need for handcrafted patterns designed by experts and refined through great effort and significant amounts of developer time (e.g., the Hex-Rays decompiler has been under active development for 18 years). In principle, this also means that neural decompilation can easily support new architectures or source languages as long as a compiler and programs in the source language are available; by contrast, Hex-Rays only supports four main CPU architectures (x86, ARM, MIPS, and PowerPC) and a single source language (C). Second, because they are trained on source code written by humans, neural decompilers will typically generate code that looks more natural and is free of artifacts like GOTO statements that plague traditional decompilers~\cite{nogoto}. Finally, particularly on modern GPUs, decompilation using NMT techniques can be much faster than a traditional decompiler.

In pursuing neural decompilation here, we take the view that neural decompilation is, for now, best used to \emph{complement} traditional decompilation. Even though a neural decompilers may not get the exact constants or operators right, they are typically good good capturing a rough high level sketch of the original function, and it may be easier for a human to fix up such a sketch by referring to the assembly or the output from a traditional decompiler. This could allow a reverse engineer to take advantage of the improved readability of NMT techniques (particularly for languages other than C) while still retaining the semantic accuracy of classical decompilers.

\subsection{Computer Languages vs. Human Languages}

\begin{figure}[htbp]
\centerline{\includegraphics[width=0.50\textwidth]{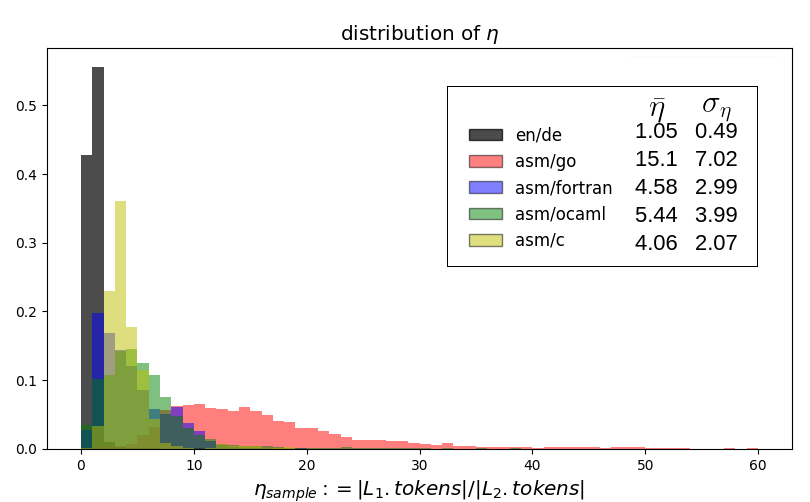}}
\caption{Programming languages are markedly different than human languages.}
\label{fig:rdist}
\end{figure}

Although we are using techniques from neural translation of human languages, programming languages are in fact very different from natural language.  To illustrate and quantify these differences, we define a parameter called \textit{sequence contraction}, denoted $\eta$: for an NMT task translating from language $L_1$ to $L_2$, the $\eta$ for a given sample is the ratio of the number of $L_1$ tokens to $L_2$ tokens. Figure~\ref{fig:rdist} shows the distribution of $\eta$ over samples from our programming languages dataset, and from the IWSLT 2012 English/German dataset~\cite{iwslt}.

This figure captures the high asymmetry between high level source code and low level assembly. Not only does the mean of $\eta$ vary greatly between human language translation and decompilation, but also between different programming languages. There is also a large difference in the standard deviation for each, suggesting that decompilation is a distinctively different domain compared to NMT for human languages.

\section{Design}

\begin{figure*}[t]
	\centering
	\includegraphics[width=\textwidth]{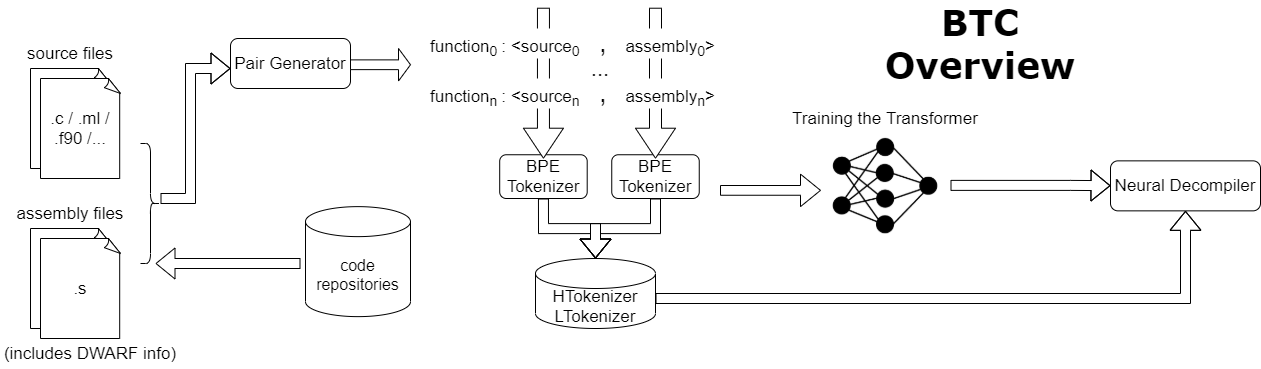}
	\caption{Overview of the system}
	\label{arc}
\end{figure*}

 In this section we discuss the high-level design of our decompiler, BTC (Beyond the C): how we prepare the training data, what model architecture we use, and how we train the model. The main goal of our design is to minimize domain knowledge and make every step as generic and as simple as possible, not depending on specifics of any programming language or CPU architecture. We formulate the problem of decompilation as a text-to-text translation, looking at both the source and machine code as a sequence of tokens and just that, with no source parsing, CPU instruction awareness or indication to the transformer about what a token \emph{means}. By eschewing domain knowledge, we hope to obtain a retargetable approach that can be applied to any language.
 
 Figure~\ref{arc} shows an overview of BTC's design. At a high level, we fetch source of programs from a code repository, compile them into assembly, and then generate pairs of assembly and source code functions from those assembly and source files. From these pairs we generate two separate tokenizers for source and assembly, respectively, and use the tokenized pairs to train our model. The final result is a trained model and tokenizers.

\subsection{Model and Data}

To train the model we need pairs of source code snippet with the corresponding assembly snippet. We generate these pairs at function level: each of our source code snippets is a whole function. The benefit of this approach is that functions are a natural decomposition of a program, which mitigates the attribution problem discussed earlier: for smaller snippets it is harder to make a correspondence between source and assembly. As an example consider a line containing a variable declaration: which assembly lines should we attribute to that line? Or what part of source code should we attribute to the function prologue and epilogue? At the function level, there is less ambiguity: we know the assembly code that was generated for the whole function.

To extract pairs of source and assembly at function level we use Call Frame Information (CFI) directives which are placed into the generated assembly by the compiler and which mark the start and end of each procedure, as well as directives indicating what source lines the body of the function were compiled from. The assembler uses these directives to put the debug information into the generated executable, usually in DWARF\footnote{https://dwarfstd.org/} format. CFI directives have a very simple format with each directive is on a separate line, making it straightforward to process the assembly (``.s'') files to extract functions. Their format is shared between both LLVM and GNU compilers, allowing us to implement the function extractor once and then use it for all four languages. Supporting additional compilers that do not use CFI directives, such as Microsoft's Visual Studio or Intel's \texttt{icc}, would require a small amount of additional effort to identify similar debug information.

Working at the function level means that this approach does not require any parsing or lexing on either source or assembly side, allowing the pair extraction process to be decoupled from the specifics of the high level language or the low level assembly. Prior work~\cite{Katz2018}, which relied on relatively short snippets, needed more sophisticated techniques to parse source and assembly code and create paired snippets.

\subsection{Preprocessing}

Once we have generated assembly files from the source files we go over the assembly files and, based on the assembler CFI directives, we extract the functions from the assembly file. In each function there are directives specifying source files and lines in that source file which correspond to the function. Before emitting the pair of assembly and source text, for each of the source lines corresponding to that function, we do some minimal preprocessing: 
\begin{itemize}
\item Normalize whitespace (tabs, spaces)
\item Remove comments
\item Replace string literals with ``STR''
\end{itemize}

Even this small amount of preprocessing is a departure from our purpose of having no language dependence, since these constructs (whitespace, comments, and string literals) may have slighting different syntax depending on the source language. We make this compromise because transformer models have quadratic ($n^2$) complexity with respect to the input sequence, so preprocessing allows us to include larger functions in our training data. Luckily, the implementation for this preprocessing is quite simple (compared to, say, building an AST) and handles the four languages in our evaluation in only 166 lines of Python.

\subsection{Tokenization}

After extracting and preprocessing our function pairs, we tokenize the snippets. We have a number of choices in how we tokenize the source and assembly text. A common choice in natural language processing (NLP) is to simply split on spaces. This results in an infeasibly large vocabulary if used on source code, however, since source code contains many unique names and individual ``words'' need not be separated by spaces (consider, e.g., ``x+2''). At the other end of the spectrum are domain specific tokenizers such as those used in prior work~\cite{Katz2018}; while these give more precise tokenization, they must be rewritten for each new source language, decreasing retargetability. We could also tokenize at the byte level, making each individual byte its own token, with a vocabulary size of 256; however, because ML models are limited in the number of tokens they can process at once, this would severely limit the size of functions we can handle. 
Our solution is to instead use byte pair encoding (BPE)~\cite{bpe}, which has become a popular choice in NLP and is used by large language models such as GPT-3. BPE is a subword tokenization based on recurring subword i-grams leading to a reasonable vocabulary size while offering some degree of compression and allowing larger functions to fit into the model's input. For the tokenizer used for the ``large'' OCaml dataset (described in Section~\ref{subsec:largeds}) represent around 5.9 characters per token on average. The vocabulary includes single character tokens like digits, keywords like OCaml's ``let" and common subword tokens like ``str\_''. Our implementation uses Huggingface's built-in BPE facility, and also splits numbers into digits (one token per digit). We provide an evaluation of byte-level vs BPE tokenization in Section~\ref{eval:tok}. 

For each language we train the tokenizer on source code and assembly code separately to obtain high-level (source) and low-level (assembly) tokenizers, which we will refer to as the HTokenizer and LTokenizer, respectively.

\section{Implementation}

\subsection{Training Data}

\begin{table*}
	\centering
	\caption{Dataset Parameters}
	\label{tbl:dataset}
	\begin{tabular}{|c|c|S[table-format=4.0]|S[table-format=5.0]|S[table-format=4.0]|S[table-format=8.0]|S[table-format=8.0]|S[table-format=9.0]|S[table-format=7.0]|}
		\hline
		Dataset & \shortstack{Max Source \\ Length} & \shortstack{Max Asm \\ Length} & \shortstack{Source Vocab \\ Size} & \shortstack{Asm Vocab\\ Size} & \shortstack{Source Line\\ Count} & \shortstack{Source Token\\ Count} & \shortstack{Asm Token\\ Count} & \shortstack{Function\\ Count}  \\
		\hline
		C-S & 271 & 1776 & 7104 & 4040  & 148178 & 1106279 & 5209251 & 12488        \\ 
		Go-S & 254 & 6350 & 8688 & 4848 & 131148 & 888912 & 15121847 & 9879         \\
		Fortran-S & 398 & 5408 & 8352 & 3056 & 127545 & 1014613 & 5758989 & 9598     \\
		OCaml-S & 192 & 5939 & 23184 & 17288 & 153695 & 1574681 & 12588749 & 70348     \\
		\hhline{|=|=|=|=|=|=|=|=|=|}
		C-L & 271 & 1776 & 11176 & 11608 & 38141177 & 319519055 & 397132136  & 2000000       \\ 
		OCaml-L & 271 & 1776 & 10960 & 11680 & 1273767 & 16651933 & 129398038 & 700000      \\ 
		\hline
	\end{tabular}
\end{table*}

One of the most important factors in the performance of a machine learning model is the quality of its training data. Here, we describe how we selected training samples for each of our four languages. For each language, we needed to build a large number of programs written in that language and obtain their assembly code; given that some languages have no standard build system, this is sometimes difficult and is the most time-consuming part of adding a language to BTC. However, our use of compiler-generated debug information allows us to reuse the same preprocessing and extraction steps for each language, as long as we can get the compiler to emit assembly in the standard format used by the GNU assembler. A summary of the collected data for each language can be seen in Table~\ref{tbl:dataset}.

\subsubsection{C-Small (C-S)}
For our small C dataset, we collected competitive programming and interview style problems (e.g., leetcode\footnote{\url{https://leetcode.com/}}, Project Euler\footnote{\url{https://projecteuler.net/}}, and UVa\footnote{\url{https://en.wikipedia.org/wiki/UVa_Online_Judge}}); we were able to find enough of these repositories to give us enough data to train and evaluate our model. These programs usually consist of a single source, are easy to compile, and typically do not use complex language features. Similar programs have also been used for corpus creation in prior neural decompilation work, allowing us to make an apples-to-apples comparison with such work.

\subsubsection{Go-Small (Go-S)}
Similar to C, for Go we also started with competitive programming and interview style repositories. Because we found fewer code samples written in Go, we augmented this with additional repositories found on GitHub, including general algorithm implementations and relatively small utility programs. As these are relatively simple programs, we were able to obtain the assembly files by simply invoking the \texttt{gccgo} compiler on each source file with the ``-S'' option.

\subsubsection{Fortran-Small (Fortran-S)}
For Fortran we picked real-world programs solving computational problems (a Couette flow solver, ODE solvers, etc.) as this is the most prevalent use case for Fortran. These repositories were the top results in a GitHub search for ``Fortran90'' and are among the most-starred Fortran repositories. For generating the assembly files from source files, we traversed each directory and attempted compilation of each source file, using several attempts to try and solve simple compilation errors due to a compilation flag or including/referencing of another source file. We used the GNU Fortran compiler (\texttt{gfortran}) for the compilation.

\subsubsection{OCaml-Small (OCaml-S)}
For OCaml there is an easy way to instruct the compiler to generate assembly files and include DWARF information by setting the \texttt{OCAMLPARAM} environment variable, which gets passed to the compiler whenever it is called, and the OPAM\footnote{\url{https://opam.ocaml.org/}} package manager has a \texttt{--keep-build-dir} flag that keeps all the intermediate files generated in the process of building a package. We built 119 packages of various sizes, including packages from Jane Street\footnote{\url{https://github.com/janestreet}} and CMU's Binary Analysis Platform (BAP)~\cite{bap}. After building packages with OPAM we traversed the build directory and parsed the ``.s" files to generate the data.

\subsubsection{Large Datasets}
\label{subsec:largeds}
In machine learning, the common wisdom holds that more data is always better. To study the effect of having more data, we also gathered two large datasets for C and OCaml; we will make these datasets publicly available to aid in future research on neural decompilation. 

For OCaml, we used the same procedure that we used for the OCaml-S dataset, but with \emph{all} packages available in OPAM. The OCaml-Large (OCaml-L) dataset consists of over 900K functions, of which 700K are short enough to be used for training. This gives us roughly 10\texttimes\ the amount of data as we had for OCaml-S.

Compiling a large number of C programs is more challenging, because there is no standardized build system used by all C projects. To solve this, we use packages included in Debian Linux, which can be automatically built in a uniform way. We created a tool that automates this process by downloading each package available in the ``main'' section of Debian Buster, and then selecting all packages that contained C source files (excluding C++) based on file extension, for a total of 50K packages. We then built these using \texttt{dpkg-buildpackage}, using Docker containers to isolate each package build and running the builds and preprocessing steps in parallel. Because the GNU C compiler (\texttt{gcc}) creates temporary assembly files for each compilation unit, we used \texttt{LD\_PRELOAD} to hook the \texttt{unlink} system call and preserve the generated assembly code for each package. On a dual-CPU AMD EPYC 7542 server with 128 parallel containers, we were able to extract 6 million functions in less than five hours, selecting 2 million for training based on sequence length.

\subsection{Training the Transformer}

For model training we use Facebook's fairseq toolkit with its standard transformer encoder/decoder architecture. The models were trained on an HPC node with an NVIDIA V100 (32 GB) or RTX8000 (48 GB) GPU with and 2x Intel Xeon Platinum 8268 CPUs with 369 GB of RAM. The model parameters (encoder layers, attention heads, etc.) can be seen in Table~\ref{tbl:modelparam}, and these same parameters were used for all the languages we trained on. The only per-language parameters are the maximum sequence lengths, which were set based on the maximum observed in the dataset for each language, and the vocabulary sizes, which can be seen in Table~\ref{tbl:dataset}. To batch the data, we used a fairseq option to specify a maximum number of tokens in a batch; fairseq then batches samples so that the number of tokens in each batch is close to that maximum value. This helped us keep GPU utilization high given that we had samples covering a wide range of token lengths.

To take advantage of the larger amount of data for C and OCaml, we also trained two larger models, with twice as many encoder and decoder layers (12 vs 6) and a larger embedding dimension (768 vs 512); however, to fit within the memory limit of the GPU we also had to reduce the number of attention heads from 16 to 8. The small models were trained until the test accuracy stopped improving, which only takes a few hours. The large models take significantly longer to train, and the C model was still (slowly) improving after ten days, at which point we terminated the training.

\begin{table}
	\centering
	\caption{Parameter and Hyperparameter values}
	\label{tbl:modelparam}
	\begin{tabular}{|c|c|c|}
		\hline
		Parameter & S-Model & L-Model            \\
		\hline
		learning rate (fixed) & 0.1   &    0.1       \\ 
		encoder layers & 6            &    12       \\
		decoder layers & 6            &    12       \\
		encoder attention heads & 16  &    8       \\
		decoder attention heads & 16  &    8       \\
		max tokens per batch & \num[mode=text]{10000}  &    \num[mode=text]{10000}       \\
		gradient clip norm & 0.1      &    0.1       \\
		embedding dimension & 512 & 768 \\
		\hline
	\end{tabular}
\end{table}

\subsection{BTC VSCode extension}

\begin{figure}[htbp]
\centering
\includegraphics[width=0.8\columnwidth]{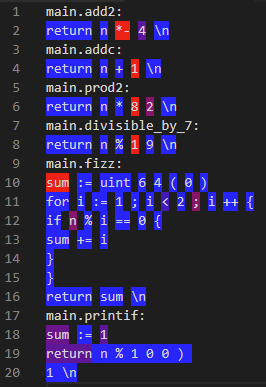}
\caption{Highlighted output of the model in BTC's VSCode extension}
\label{vsx}
\end{figure}

To observe the results for qualitative assessment and allow interactive exploration, we also created a Visual Studio Code (VSCode) extension (shown in Figure~\ref{vsx}) that can be called on any ``.s" file and will run every procedure in that assembly file through the model, showing the model's decompilation with each token colored according to its probability according to the model. Such a view could also help a human agent to fix the errors based on seeing which tokens have lower confidence scores. The extension consists of 146 lines of TypeScript code that extract the functions in an open ``.s" file, run the decompilation model using an external Python script (100 lines), and then render the results in a side panel within VSCode; we will also release the extension.

\section{Evaluation}

\begin{table}
	\centering
	\caption{Summary of evaluation results}
	\label{rez}
	\begin{tabular}{|c|S[table-format=1.2]|S[table-format=2.2]|S[table-format=9.0]|}
		\hline
		\makecell{Model} & \shortstack{Avg.\\ Edit Dist.} & \shortstack{Translation Speed\\ (function/s)} & \shortstack{Number of\\ Model Params} \\ 
		\hline
		C-S-S        & 0.60            & 17.8   & 53481472 \\ 
		Go-S-S       & 0.74            & 14.74  & 55517184 \\ 
		Fortran-S-S  & 0.63            & 13.4   & 54255616 \\ 
		OCaml-S-S    & 0.77            & 27.7   & 76730368 \\ 
		\hline
		C-L-S     &   0.55     &   18.4    &    61747200    \\
		C-L-L     &   0.46     &   11.52   &    187115520    \\
		OCaml-L-S   &   0.59     &   20.81   &    61710336    \\
		OCaml-L-L   &   0.60     &   18.2   &    187060224    \\
		\hline
		Katz et al.~\cite{Katz2018} & 0.70 & N/A & N/A \\
		\hline
	\end{tabular}
\end{table}

\subsection{Quantitative Metric}

We measure decompilation quality using (normalized) average edit distance (AED), the average of edit distances between prediction and ground truth, divided by length of the ground truth sequence ; this metric is also used by Katz et al.~\cite{Katz2018}, the most closely related prior work to BTC. In the translation of human languages, measures like BLEU~\cite{bleu} are also common, but prior work has found that BLEU may not be appropriate for code~\cite{Tran}. Average edit distance also intuitively captures how we believe BTC might be used, as it roughly indicates how many changes a reverse engineer might have to make to the decompiled function in order to recover the original code (relative to the size of the function).

The AED for samples in our held-out test set for each model are shown in Table~\ref{rez}. For comparison, Katz et al.~\cite{Katz2018} achieved an average edit distance of 0.70 over snippets which were an order of magnitude shorter than our samples. However, in absolute terms, even our best result (for the large C model trained on 2M functions, C-L-L) is still relatively poor: an AED of 0.46 means (very roughly) that changes are needed to half the tokens in a given function. In practice, we have found that this is not quite true; many functions are decompiled almost perfectly, while others are completely wrong. But it is clear that neural decompilation is not yet ready for practical use on real code.

\subsection{Effect of Dataset and Model Size}

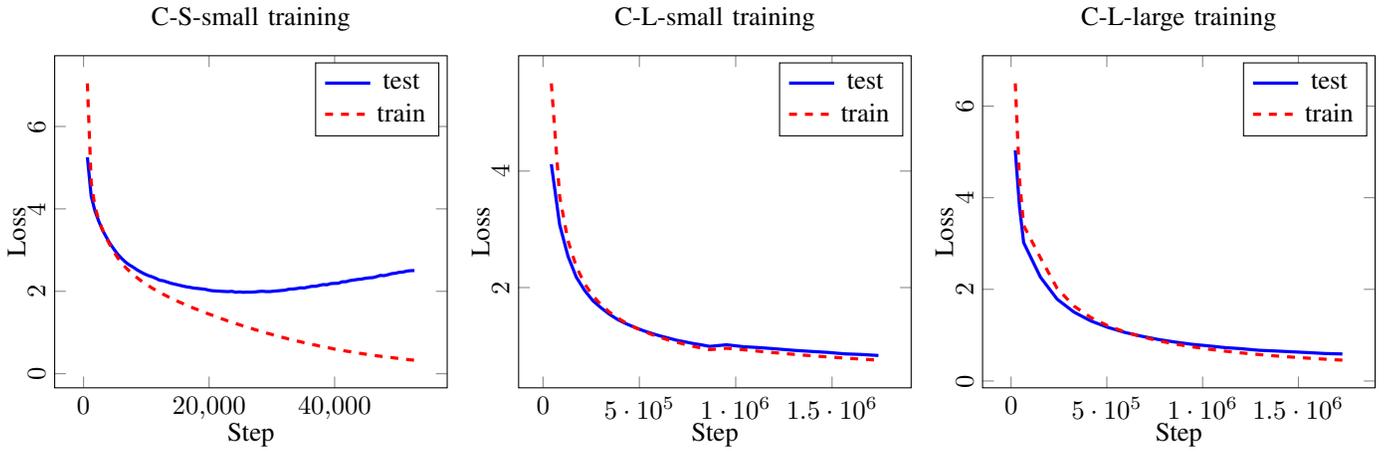
\begin{figure*}
\begin{minipage}[t]{0.31\textwidth}
\begin{tikzpicture}
\begin{axis}[title=C-S-small training,
scaled x ticks = false,
width=6.8cm,height=6cm,xlabel=Step,
every axis y label/.style={at={(-0.05,0.38)},above=4mm,rotate=90},
every axis x label/.style={at={(0.5,0)},above=-9mm},
y tick label style={rotate=90},
ylabel=Loss, no markers,
every axis plot/.append style={very thick}]
\addplot table [mark=none,x=Step, y=Value, col sep=comma] {CD_test.csv};
\addplot [dashed, red] table [mark=none,x=Step, y=Value, col sep=comma] {CD_train.csv};
\legend{test,train}
\end{axis}
\end{tikzpicture}
\end{minipage}
\hspace{0.3cm}
\begin{minipage}[t]{0.31\textwidth}
\begin{tikzpicture}
\begin{axis}[title=C-L-small training,
scaled x ticks = false,
width=6.8cm,height=6cm,xlabel=Step,
every axis y label/.style={at={(-0.05,0.38)},above=4mm,rotate=90},
every axis x label/.style={at={(0.5,0)},above=-9mm},
y tick label style={rotate=90},
ylabel=Loss, no markers,
every axis plot/.append style={very thick}]
\addplot table [mark=none,x=Step, y=Value, col sep=comma] {CS_test2.csv};
\addplot [dashed, red] table [mark=none,x=Step, y=Value, col sep=comma] {CS_train2.csv};
\legend{test,train}
\end{axis}
\end{tikzpicture}
\end{minipage}
\hspace{0.3cm}
\begin{minipage}[t]{0.31\textwidth}
\begin{tikzpicture}
\begin{axis}[title=C-L-large training,
scaled x ticks = false,
width=6.8cm,height=6cm,xlabel=Step,
every axis y label/.style={at={(-0.05,0.38)},above=4mm,rotate=90},
every axis x label/.style={at={(0.5,0)},above=-9mm},
y tick label style={rotate=90},
ylabel=Loss, no markers,
every axis plot/.append style={very thick}]
\addplot table [mark=none,x=Step, y=Value, col sep=comma] {CL_test.csv};
\addplot [dashed, red] table [mark=none,x=Step, y=Value, col sep=comma] {CL_train.csv};
\legend{test,train}
\end{axis}
\end{tikzpicture}
\end{minipage}\caption{Comparison of training loss-step for C-S-small, C-L-small and C-L-large}
\label{fig:ctrain}
\end{figure*}

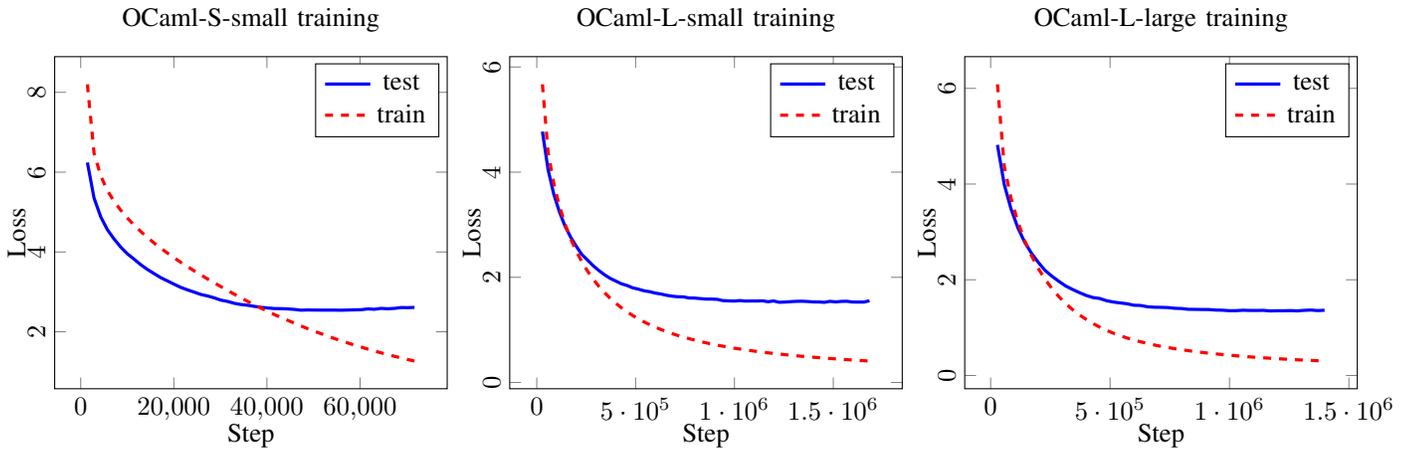
\begin{figure*}
\begin{minipage}[t]{0.13\textwidth}
\begin{tikzpicture}
\begin{axis}[title=OCaml-S-small training,
scaled x ticks = false,
width=6.8cm,height=6cm,xlabel=Step,
every axis y label/.style={at={(-0.05,0.38)},above=4mm,rotate=90},
every axis x label/.style={at={(0.5,0)},above=-9mm},
y tick label style={rotate=90},
ylabel=Loss, no markers,
every axis plot/.append style={very thick}]
\addplot table [mark=none,x=Step, y=Value, col sep=comma] {OD_test.csv};
\addplot [dashed, red] table [mark=none,x=Step, y=Value, col sep=comma] {OD_train.csv};
\legend{test,train}
\end{axis}
\end{tikzpicture}
\end{minipage}
\hspace{0.19\linewidth}
\begin{minipage}[t]{0.13\textwidth}
\begin{tikzpicture}
\begin{axis}[title=OCaml-L-small training,
scaled x ticks = false,
width=6.8cm,height=6cm,xlabel=Step,
every axis y label/.style={at={(-0.05,0.38)},above=4mm,rotate=90},
every axis x label/.style={at={(0.5,0)},above=-9mm},
y tick label style={rotate=90},
ylabel=Loss, no markers,
every axis plot/.append style={very thick}]
\addplot table [mark=none,x=Step, y=Value, col sep=comma] {OS_test.csv};
\addplot [dashed, red] table [mark=none,x=Step, y=Value, col sep=comma] {OS_train.csv};
\legend{test,train}
\end{axis}
\end{tikzpicture}
\end{minipage}
\hspace{0.19\linewidth}
\begin{minipage}[t]{0.13\textwidth}
\begin{tikzpicture}
\begin{axis}[title=OCaml-L-large training,
scaled x ticks = false,
width=6.8cm,height=6cm,xlabel=Step,
every axis y label/.style={at={(-0.05,0.38)},above=4mm,rotate=90},
every axis x label/.style={at={(0.5,0)},above=-9mm},
y tick label style={rotate=90},
ylabel=Loss, no markers,
every axis plot/.append style={very thick}]
\addplot table [mark=none,x=Step, y=Value, col sep=comma] {OL_test.csv};
\addplot [dashed, red] table [mark=none,x=Step, y=Value, col sep=comma] {OL_train.csv};
\legend{test,train}
\end{axis}
\end{tikzpicture}
\end{minipage}
\caption{Comparison of training loss-step for OCaml-S-small, OCaml-L-small and OCaml-L-large}
\label{fig:ocamltrain}
\end{figure*}

Training on larger dataset enhanced the performance of the model for both C and OCaml. Figures~\ref{fig:ctrain} and \ref{fig:ocamltrain} show how overfitting is mitigated with larger datasets, as both training and test loss converge together as opposed to test loss starting to rise or level off before training loss. However, there are still limits to how good the accuracy gets, as the very large 2M C dataset shows. And although with larger datasets a bigger model can improve results, we only observed this with the 2M C dataset; the larger model shows only a minor improvement on the 700K OCaml dataset.

\subsection{Translation Performance}

The other metric of concern is translation (inference) time, also shown in Table~\ref{rez}. All our models translate more than 10 functions per second on average on a V100 GPU, which is quite fast. However, we note that this is not directly comparable to a traditional decompiler like GHIDRA or Hex-Rays, as these tools also have to do the work of disassembly, control flow recovery, etc., whereas we assume that these tasks have already been done and that the input assembly is in a form comparable to what the compiler produced. That being said, the relatively fast translation speed means that there is room to pair BTC with more costly techniques that could fix up our candidate sketches.

\subsection{Tokenization}
\label{eval:tok}

To validate our choice of tokenization, we compared BPE tokenization to character-level tokenization on our Fortran dataset with the small model configuration. Because the character-level encoding is less efficient, we limited the training dataset for the character-level (CL) model to functions less than 6{,}000 tokens (throwing away ~30\% of the dataset). The BPE model is the Fortran-S model trained on the full Fortran dataset. To make a direct comparison, the models were evaluated on the reduced test set and the AED was computed by tokenizing the CL model's output using the BPE model's HTokenizer. The two models performed about equally well: the CL model achieved an AED of 0.59 while the BPE model scored 0.62.

We note that the reduced test set, which consists of smaller functions, is also likely to be somewhat easier to decompile, so the small improvement seen here is not likely to be significant. Because of this, and the fact that the CL model cannot support larger functions, we believe that the use of BPE is justified.

\begin{figure*} 
Fortran: 0.26 \sout{\hfill} \\
\begin{minipage}[b]{0.45\textwidth}
\centering
\begin{minted}[breaklines]{Fortran}
subroutine en_her_02_xiu_size ( n , o ) 
 implicit none 
 integer ( kind = 4 ) n 
 integer ( kind = 4 ) o 
 o = n + 1 
 return 
end
\end{minted}
\end{minipage}
\hspace{0.5cm}
\begin{minipage}[b]{0.45\textwidth}
\centering
\begin{minted}[breaklines]{Fortran}
subroutine i4_determinant ( n , value ) 
 implicit none 
 integer ( kind = 4 ) n 
 integer ( kind = 4 ) value 
 value = n * n - 1 
 return 
end
\end{minted}
\end{minipage} \\
OCaml: AED 0.62 \sout{\hfill} \\

\begin{minipage}[t]{0.45\textwidth}
\centering
\begin{minted}[breaklines]{OCaml}
let fmt_path f x = fprintf f " STR " fmt_path_aux x ;;
\end{minted}
\end{minipage}
\hspace{0.5cm}
\begin{minipage}[t]{0.45\textwidth}

\centering
\begin{minted}[breaklines]{OCaml}
let pp ppf x = Format . fprintf ppf " STR " ( to_string x )
\end{minted}
\end{minipage} \\
Go: 0.14 \sout{\hfill} \\
\begin{minipage}[t]{0.45\textwidth}

\centering
\begin{minted}[breaklines]{Go}
sum := 0 
for _ , val := range arr { 
sum += val
} 
return sum
\end{minted}
\end{minipage}
\hspace{0.5cm}
\begin{minipage}[t]{0.45\textwidth}
\centering
\begin{minted}[breaklines]{Go}
sum := 0 
for _ , v := range nums { 
sum += v
} 
return sum
\end{minted}
\end{minipage} \\
C: 0.84 \sout{\hfill} \\
\begin{minipage}[t]{0.45\textwidth}

\centering
\begin{minted}[breaklines]{C}
char* p = "STR";
while ( scanf ("STR", &a )== 1 && a ) puts ( p );
return 0; 
\end{minted}
\end{minipage}
\hspace{0.5cm}
\begin{minipage}[t]{0.45\textwidth}
\centering
\begin{minted}[breaklines]{C}
while (scanf ("STR", &a) != EOF) if (a == 0) 
printf ("STR");
else
printf ("STR");
return 0;
\end{minted}
\end{minipage}
\caption{Example BTC decompilations and normalized edit distance from ground truth - left: ground truth, right: prediction}
\label{ql}
\end{figure*}

\subsection{Qualitative Observations}

To get a better feel for the decompilations produced by BTC, we have show in Figure~\ref{ql} one sample decompilation from each language alongside its ground truth; samples were chosen by hand to illustrate strengths and weaknesses of the decompiler. All samples shown were produced by the small models.

Looking at some of the examples, we observed that on all four models, the general structure is picked up fairly well. Major features like loops and conditionals correspond fairly well to the original code. The variable names even make sense in many instances, which illustrates a potential advantage of the neural approach (one explored in more detail in Jaffe et al.~\cite{jaffe2018}).

However, the models often make small but semantically meaningful mistakes: in the Fortran example, ``n + 1'' becomes ``n * n + 1''. Such mistakes could potentially be fixed up by a human by comparing the decompiled output to the assembly or the output of a traditional decompiler, which may still be easier than trying to manually translate the C output of a traditional decompiler into Fortran. The quality also varies between different languages; OCaml appears to be the hardest to decompile of the four languages we investigated.

We also see that edit distance does not perfectly capture decompilation quality. In the C example, BTC is penalized for using printf instead of puts, for not storing the string in a separate variable (p), and for checking for EOF rather than 1 as the return value from scanf. These choices are arguably semantically equivalent to the original code in this case, but are completely different in terms of edit distance.

Still, neural decompilers have a long way to go. The decompiled output is not nearly accurate enough to be trusted without additional manual or automated fixups. However, we believe that these results do indicate that neural decompilation remains a promising research area, and we hope that our work can help establish a baseline and proof of concept of a multi-language approach to decompilation.

\subsection{Quality of Training Data}
\label{dataquality}

One major challenge of using code that is used in the wild for training data is that unlike simple synthetic data there is a great variety in the structures and idioms used and we realized there are many issues that can adversely affect the quality of training data. Our function extractor uses ``.loc" directives in assembly to attribute source lines to assembly procedures. In a language with a simple structure like Fortran, each function will be represented in the source code in a contiguous set of lines, and so the loc directives can be used to match to that region.

In more complex languages, however, this mapping may not be straightforward. For example, we may we have anonymous functions which perforate the otherwise connected regions, the effect of such problems with matching source code and assembly also depend on how frequent these constructs happen. OCaml turned out to be the most problematic with lowest quality of training data; upon inspection, many of the training examples are not actually complete functions (or even syntactically valid due to mistakes in extraction). This likely explains the poor performance of the decompiler on OCaml, even with a relatively large training dataset and model.

These problems are also likely to carry over to other functional languages like Haskell, in which functions are first class citizens and lambdas are used frequently. However, even C is prone to such issues, as features like preprocessor macros can break the simple mapping between source and assembly represented in the debug information.

Neural Decompilation has not been previously attempted with any language except C, and it is clear that adapting it to tackle new languages will be more difficult for some languages more than others. Our four attempts cover a variety of different language paradigms, and we hope that the dataset can be useful for future research in how language and compiler design affect training data collection. We also believe it may provide motivation for improving the quality of compiler-emitted debug information, to allow more precise matching of source constructs to the generated assembly.

\section{Related Work}

Reverse engineering of binary code has been the subject of many research works \cite{phoenix,cifuentes,johnny}. Research \cite{phoenix,retdec}, commercial (HexRays) and open source (GHIDRA) decompilers have been developed targeting the C language using the traditional approach. There have been attempts to make traditional methods produce more human-readable code and construct more structured control flow (e.g. fewer GOTOs) \cite{Durfina2013,nogoto,Chen2010,noteasy}. Durfina et al.~\cite{Durfina2011} proposed a method to automate making decompilers by using a formal specification of the hardware platform. An important phase in traditional decompilers is lifting the binary to an intermediate representation. Hasabnis et. al \cite{hasabnis} developed a learning-based method to automate this, which depended heavily on integration with the compiler. Wang et al.~\cite{ouroboros} created Ouroboros, a tool to disassemble binaries in such a way that the output can be re-assembled into a binary again which is rule-based as opposed to learning-based. Pei et al~\cite{xda} made XDA which is a learning-based disassembler.

There has also been research on specific sub-tasks, such as detecting functions inside a binary using neural networks~\cite{recog} or suggesting sensible variable names~\cite{recovername} or type annotations \cite{bigcode}, the latter two targeting Javascript. Representation learning methods have been applied to assembly code to encode instructions as vectors~\cite{palmtree,asm2vec,peng} or infer variable types~\cite{stateformer}, which are useful in downstream tasks like finding semantically similar code. Another approach for this task is Trex~\cite{trex}, which uses transfer learning to learn execution semantics from execution traces.

In neural decompilation, there have been several prior efforts~\cite{Katz2018,coda,nbref}, all  of which focused on the C language. We also note that this work focused almost entirely on very small programs derived from sources like programming challenges, which does not reflect the kind of code that is usually of interest to reverse engineers. We believe that our work, particularly the larger C dataset, represents the largest test of neural decompilation on real-world code to date.

\section{Limitations and Future Work}

Traditional decompilers operate with a completely different approach which consists of multiple stages of parsing and usually intermediate-languages like GHIDRA's pcode. They rely on hardcoded idioms, and data-flow and type analysis tailored for C. Although they can generate hard-to-understand code, traditional decompilers have the advantage that unlike neural networks, they are interpretable. Interpretability is a major drawback in deep learning methods and has been an active field of research\cite{interp}.

Given the black-box nature of deep networks, these methods cannot guarantee semantic correctness and suffer from lower semantic accuracy. However, there are techniques that can enhance the accuracy of neural decompilers. Schulte et al~\cite{Schulte2018} use evolutionary algorithms to mutate a code sketch and compile it until the generated object code is byte-equivalent with the input. Our approach can be used for the sketch generation of their method. Another approach is to delegate some error handling to the human agent, as we suggest in the Evaluation section: many of the errors are the network getting a constant or operator wrong, which can be readily resolved by a human agent with access to the assembly code or other tools. Advanced data normalization, like the canonicalization scheme described by Katz et al.~\cite{Katz2019TowardsND} can also enhance accuracy if it can be automated. 

A limitation specific to BTC is dependence on the debug info, the CFI directives emitted by the compiler to let the assembler generate DWARF information. The quality of implementation of these information and DWARF support varies across different compilers and may even be buggy. As we note in Section~\ref{dataquality}, progress in neural decompilation may depend on improving the quality of the source to assembly mapping provided by compilers.
 
Finally, all NMT methods suffer from limitations on length of input sequences supported. Previous methods based on RNNs were severely limited and could only handle small snippets of code, this might be one of the reasons that no work had tackled any language besides C until now, as Figure \ref{fig:rdist} shows, other languages produce more verbose assembly code that would exacerbate this problem, shorter snippets also suffer more from the attribution problem and would make it harder for the model to learn larger constructs. We predict that this limitation will be alleviated as NMT methods in general and learning hardware gets more advanced. They have been advancing between RNNs and the more recent Transformer models, compared to the previous work~\cite{Katz2018} we managed to expanded the size limit by an order of magnitude.

\section{Conclusion}

In this paper we demonstrated that neural decompilation can indeed provide a path to retargetable decompilers for different programming languages. In addition to C, we report results on applying neural decompilation to Fortran, Go and OCaml, which have not previously been evaluated. Our approach, Beyond The C (BTC), does not rely on compiler customization, complex parsing, or even specifying keywords or operators of the programming language, which eases the burden of supporting new languages while matching the quality of comparable prior work; despite the lack of any explicit knowledge about the source languages or CPU architectures baked in, we saw that the model can replicate many of the features of the language. Although many challenges remain, particularly related to code size limitations, semantic accuracy, and data quality, we hope that other researchers will see our work as a sign that there is fertile ground for new research beyond the C.

\section*{Availability}

Our code, data and trained models can be found at:

\url{https://figshare.com/s/2c68b9c181e80f4e3b06}

\section*{Acknowledgments}
The authors would like to thank Ivan Gotovchits for helpful suggestions about working with the OCaml build system. This research was supported in part by National Science Foundation (NSF) Award 1801495. Any opinions, findings, conclusions, or recommendations expressed are those of the authors and not necessarily of the NSF.

\bibliographystyle{plain}
\bibliography{paper}
\vspace{12pt}

\end{document}